\documentclass[fleqn,usenatbib]{mnras}

\usepackage{newtxtext,newtxmath}

\usepackage[T1]{fontenc}
\usepackage{ae,aecompl}

\usepackage{graphicx}	
\usepackage{amsmath}	
\usepackage{amssymb}	

\usepackage{abbrevs}
\usepackage{epstopdf}

\title[The largest glitch observed in the Crab pulsar]{The largest glitch observed in the Crab pulsar}

\author[B. Shaw et al.]{B. Shaw$^{1}$\thanks{E-mail: benjamin.shaw@manchester.ac.uk},
A.G. Lyne$^{1}$,
B.W. Stappers$^{1}$,
P. Weltevrede$^{1}$,
C.G. Bassa$^{2}$,
A.Y. Lien$^{3,4}$,
\newauthor M.B. Mickaliger$^{1}$,
R.P. Breton$^{1}$,
C.A. Jordan$^{1}$,
M.J. Keith$^{1}$
and H.A. Krimm$^{5,6}$
\\
$^{1}$Jodrell Bank Centre for Astrophysics, School of Physics and Astronomy, University of Manchester, Manchester, UK, M13 9PL\\
$^{2}$ASTRON, the Netherlands Institute for Radio Astronomy, Postbus 2, NL-7990 AA Dwingeloo, the Netherlands\\
$^{3}$Center for Research and Exploration in Space Science and Technology (CRESST) and NASA Goddard Space Flight Center, \\ ~Greenbelt, MD 20771, USA \\
$^{4}$Department of Physics, University of Maryland, Baltimore County, 1000 Hilltop Circle, Baltimore, MD 21250, USA \\
$^{5}$Universities Space Research Association, 7178 Columbia Gateway Dr, Columbia, MD 21046 \\
$^{6}$National Science Foundation, 2415 Eisenhower Ave, Alexandria, VA 22314 \\
}

\date{Accepted XXX. Received YYY; in original form ZZZ}

\pubyear{2018}

\begin{document}
\label{firstpage}
\pagerange{\pageref{firstpage}--\pageref{lastpage}}
\maketitle

\begin{abstract}
We have observed a large glitch in the Crab pulsar (PSR B0531+21). The glitch occurred around MJD 58064 (2017 November 8) when {the pulsar underwent an increase in the rotation rate of $\Delta \nu = 1.530 \times 10^{-5}$ Hz, corresponding to a fractional increase of $\Delta \nu / \nu = 0.516 \times 10^{-6}$} making this event the largest glitch ever observed in this source. Due to our high-cadence and long-dwell time observations of the Crab pulsar we are able to partially resolve a fraction of the total spin-up of the star. This delayed spin-up occurred over a timescale of $\sim$1.7 days and is similar to the behaviour seen in the 1989 and 1996 large Crab pulsar glitches. The spin-down rate also increased at the glitch epoch by $\Delta \dot{\nu} / \dot{\nu} = 7 \times 10^{-3}$. In addition to being the largest such event observed in the Crab, the glitch occurred after the longest period of glitch inactivity since at least 1984 and we discuss a possible relationship between glitch size and waiting time. No changes to the shape of the pulse profile were observed near the glitch epoch at 610 MHz or 1520 MHz, nor did we identify any changes in the X-ray flux from the pulsar. The long-term recovery from the glitch continues to progress as $\dot{\nu}$ slowly rises towards pre-glitch values. In line with other large Crab glitches, we expect there to be a persistent change to $\dot{\nu}$. We continue to monitor the long-term recovery with frequent, high quality observations.  
\end{abstract}




\section{Introduction}
PSR B0531+21 (known commonly as the Crab pulsar, due to its position at the centre of the Crab Supernova remnant, SN 1054) is one of the most widely studied members of the pulsar population, exhibiting emission across the entire electromagnetic spectrum.  On MJD 58058, the pulsar had a spin frequency ($\nu$) of 29.6Hz and a spin-frequency derivative ($\dot{\nu}$) of $-3.7 \times 10^{-10}$ Hz s\textsuperscript{-1}.  It is one of the youngest pulsars known, having formed in the supernova of \AD 1054. 




In general, pulsar spin-down is smooth and continuous however the rotation of young pulsars, such as the Crab, is occasionally interrupted by discrete spin-up events known as glitches (see \cite{elsk11} and \cite{ymhj+13} for comprehensive reviews of pulsar glitches). These are characterised by a sudden increase to the rotation rate of the star ($\Delta \nu$).  Typically the young pulsar relaxes back towards the pre-glitch spin frequency after a few days. In addition to $\Delta \nu$, the spin-down rate of the pulsar is also observed to increase and relaxes back towards pre-glitch values over many hundreds of days.  Often the occurrence of subsequent glitches interrupts the recovery resulting in glitches having a cumulative effect on long-term spin-down (e.g. \citealt{lpgc96}).  Occasionally, glitching is seen in older pulsars. In these cases, the pulsar generally exhibits minimal or no recovery towards the pre-glitch spin-frequency and changes to $\dot{\nu}$ are not generally observed (e.g. \citealt{sl96}).


Discovered in 1968 \citep{sr68}, the Crab pulsar has been observed daily at 610 MHz by the 42-ft telescope at Jodrell Bank Observatory since 1984, providing a high cadence, long baseline dataset from which to study its rotational behaviour \citep{ljg+15}.  As of October 2017, the Crab pulsar has been seen to exhibit 24 glitches \citep{ljg+15}. These glitches occur at intervals of several years with no clear periodicity between them. The magnitude of the spin-frequency increase is usually of the order $10^{-9} < \Delta \nu < 10^{-7}$ Hz. Glitches in the Crab often exhibit changes to the spin-down rate that do not fully relax back to pre-glitch values, thereby having a long term cumulative effect on the pulsar's rotation.  During the period 1995-2009 the pulsar exhibited particularly high glitch activity making the long term study of post-glitch rotation problematic due to interruption by subsequent glitches. The study of three large, relatively isolated glitches, however showed that after the initial increase in $\dot{\nu}$, the pulsar rapidly (over $\sim$100 days) recovers a fraction of it's pre-glitch $\dot{\nu}$. From then on, the pulsar resumes an increase in $\dot{\nu}$ that can be described by an exponential that persists for 2-3 years after the event \citep{ljg+15}.  

The physical mechanism driving glitch events is not yet fully resolved. Glitches are commonly understood to arise from a two-component model of neutron star structure in which a differential rotation develops between the crust and the neutron superfluid interior \citep{aaps81}.  The superfluid interior rotates due to the presence of quantised vortices which are pinned to inner crust lattice sites. Occasionally these vortices become unpinned and a coupling is induced between the interior fluid and the crust. This coupling forces the crust to spin-up resulting in the pulsar being observed to glitch (see \cite{hm15} for a recent review of glitch models). The coupling between the inner fluid and the crust is predicted to take place on timescales of less than one minute (e.g., \citealt{als84}).  Observations of glitches therefore, offer an opportunity to probe the nuclear physics of neutron star interiors and together with future measurements of neutron star radii, will constrain equations-of-state of ultradense matter.


In this work we report the Crab's 25th observed glitch which occurred on the 2017 November 8 (MJD 58064). The fractional change in the spin-rate was preliminarily measured to be $\Delta \nu / \nu = 0.471 \times 10^{-6}$ (\citealt{slb+17}; \citealt{kjbm17}) making this the largest glitch that the Crab has ever been observed to exhibit - more than twice the size of the previous largest.  In terms of its amplitude, this glitch is more akin to those consistently seen in the Vela pulsar than any glitch the Crab has previously undergone. The structure of this paper is as follows: The observations and data are described \S2, and we present our analysis in \S3. We discuss the implications of our results and make concluding remarks in \S4 and \S5. 

\section{Observations}

42-ft telescope observations are made using a bandwidth of 5-10 MHz centred on 610 MHz. The time resolution for the observations used here is 0.0330 ms. There are two observations per day with approximate durations of 9 and 3 hours.  The data are recorded in 1 minute subintegrations formed from the addition of all pulses which were observed during that time. These data are supplemented by less frequent observations with the 76-m Lovell telescope, also at Jodrell Bank. The Lovell data used here are obtained  using a bandwidth of 384 MHz centred on 1520 MHz. Radio frequency interference is mitigated using a median-filtering algorithm as well as manual inspection of individual frequency channels and subintegrations. The cleaned data are integrated over frequency and time to form a single pulse profile for each observation. 

Pulse times-of-arrival (TOAs) are formed from the cross-correlation of each observation with a high signal-to-noise (S/N) template profile.  The local TOA of the pulse (SAT) is computed by reference to local time standards. SATS are then transferred to the Solar system barycentre with the JPL DE200 ephemeris \citep{sta82}. TOAs are analysed in PSRTIME\footnote{\url{http://www.jb.man.ac.uk/pulsar/observing/progs/psrtime.html}} using a model containing the pulsar's spin-frequency (${\nu}$) and its first two time-derivatives, its position and dispersion measure parameters.  

\section{The glitch of 2017 November 8}
\subsection{Timing analysis}

The difference between the actual TOAs and those predicted by a timing model (the timing residuals) should form a normal distribution about zero if the pulsar's rotation and emission are well described by the model.  The physical effect of a pulsar glitch is to increase the spin-frequency of the star causing pulses to arrive progressively earlier than the timing model predicts. This manifests as timing residuals becoming increasingly negative with time.

\begin{figure}
   \includegraphics[width=1.0\columnwidth]{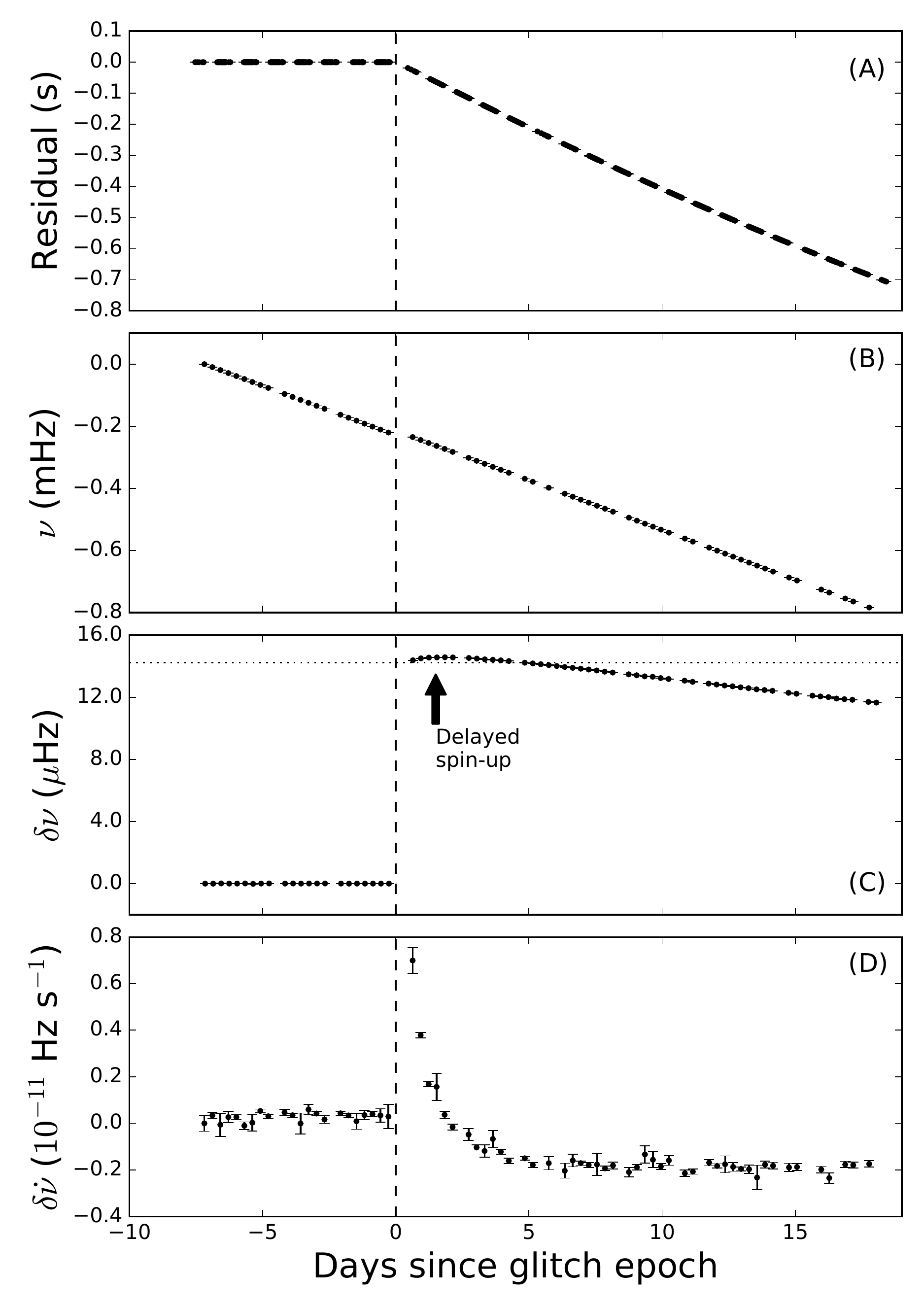}
   \caption{Panel A: Timing residuals relative to a simple pre-glitch spin-down model (see text) over 25 days consisting of $\sim$10 days prior to the glitch and $\sim$15 days after it. A fit for $\nu$ and $\dot{\nu}$ for the first 15 days of the data yields a flat distribution of residuals with values close to zero. From the glitch epoch (vertical dashed line) onwards, the pulses arrive progressively earlier than expected by the pre-glitch model. Panel B: The evolution of the spin-frequency ${\nu}$ with time over the glitch period relative to the value at the earliest data point. The glitch can be seen as a small discontinuity in the slope near the dashed line. Panel C: The {spin-frequency residuals $\delta \nu$} relative to the extrapolation of the pre-glitch rate after subtraction of the pre-glitch spin-down. The dotted line shows the amplitude of the initial rise in $\nu$.  Panel D: The evolution of $\dot{\nu}$ with time relative to an initial value of $\dot{\nu}$ (see Table 1). A downward deflection represents an increase in $|\dot{\nu}|$.  The data in the lower three panels was computed using a striding boxcar (see text).} 
   \label{spin}
\end{figure}

On the 2017 November 8, the Crab pulsar suffered a large glitch while the pulsar was below the horizon at Jodrell Bank.  In order to measure the short term rotation of the pulsar close in time to the glitch, we split the 9 and 3 hours long daily observations  into 422 individual 30 minute long observations over the time period $58057 < \textrm{MJD} < 58082$.  We show the corresponding timing residuals relative to a simple spin-down model in Figure \ref{spin} (top panel).  These parameters are listed in Table \ref{glitch_table}. Prior to the glitch (dashed vertical line), pulses were arriving as predicted by this model. After this, a significant negative gradient is seen showing that pulsar's rotational behaviour has undergone a discontinuous change and a new model is needed.  The glitch epoch is determined by generating a pair of timing models that describe the rotation immediately before and after the glitch. The two models share a common epoch that is set to be close in time to the glitch and each makes a distinct prediction of future and past TOAs. The intersection of these two models is taken to be the glitch epoch.  We compute this to be MJD 58064.555(3). The bracketed quantity is the $1\sigma$ uncertainty on the last quoted digit. The onset of the glitch occurred approximately 5 hours after the end of the last pre-glitch observation and approximately 6 hours prior to the start of the following observation.  The values for the step changes in $\nu$ and $\dot{\nu}$ are determined by measuring the difference between these values as predicted by each model at the glitch epoch.  We find the fractional change in spin-frequency to be $\Delta \nu / \nu = 0.51637(10) \times 10^{-6}$. 

In order to understand the evolution of the $\nu$ and $\dot{\nu}$ over time near the glitch we employ a striding boxcar to fit small, consecutively overlapping segments of data.  Each boxcar has a width of 1.5 days within which is contained individual TOAs from 30 minute long subintegrations. The boxcar strides over the data in steps of 0.3 days and a fit for $\nu$, $\dot{\nu}$ and $\ddot{\nu}$ is performed at each stride. The MJD of each fit is set to be halfway between the start and end of the boxcar.  The evolution of the $\nu$ and $\dot{\nu}$ over 25 days near the glitch are shown in the lower three panels of Figure \ref{spin}.


{In panel B, a comparison between the projected $\nu$ based on the pre-glitch data and the measured post-glitch $\nu$ indicates that there has been a `step' in $\nu$.} It is difficult to recognise from panel B, what the effect of the glitch on the spin-down rate (manifesting as a change in the gradient of $\nu$ with time) is.  To resolve this, in panel C we plot the change in $\nu$ having subtracted the pre-glitch $\dot{\nu}$ (Table \ref{glitch_table}) from the data presented in panel B. The glitch bears a remarkable resemblance to the Crab glitch of 1989 \citep{lsp92} as the initial spin-up of the glitch comprises two components - an initial unresolved spin-up, denoted by the dotted horizontal line in Figure \ref{spin}, followed by a resolved \emph{delayed} spin-up, in which the value of $\nu$ continues to rise for a short time. Following the delayed spin-up, the gradient has a clear negative value, indicating that the spin-down rate has increased. The evolution of the spin-down rate is plotted in panel D and shows clearly the unresolved spin-up, manifested as a rapid increase in $\delta \dot{\nu}$. The delayed spin-up is reflected in the exponential downward inflection in spin-down. 

\begin{center}
\begin{table}

    \caption{Table of pulsar/glitch parameters.}
    \begin{tabular}{ | p{3.5cm} | p{3.5cm} | }
    \hline
    Parameter & Value \\ \hline
    Period/DM epoch (MJD) & 58058.01137 \\
    Initial $\nu$ & 29.6369248116(4) Hz  \\
    Initial $\dot{\nu}$ & $-3.686703(22) \times 10^{-10}$ Hz s\textsuperscript{-1} \\
    Initial $\ddot{\nu}$ & $1.91(5) \times 10^{-19}$ Hz s\textsuperscript{-2} \\
    DM & $56.75847(31)$ pc cm\textsuperscript{-3} \\
    DM/dt & -0.028(7) pc cm\textsuperscript{-3} s\textsuperscript{-1} \\
    Glitch epoch (MJD) & 58064.555(3) \\
    Unresolved spin-up $\Delta \nu$ & $1.4233(5) \times 10^{-5}$ Hz \\
    Delayed spin-up $\Delta \nu_d$ & $1.071(4) \times 10^{-6}$ Hz \\
    $ \Delta \nu_d$ time constant $\tau_d$  & 1.703(13) days \\
    $\Delta \dot{\nu}$ & $-2.569(8) \times 10^{-12}$ Hz s\textsuperscript{-1}  \\
    \hline
    \end{tabular}
    \label{glitch_table}
\end{table}
\end{center}

\begin{figure}
    \includegraphics[width=1.0\columnwidth]{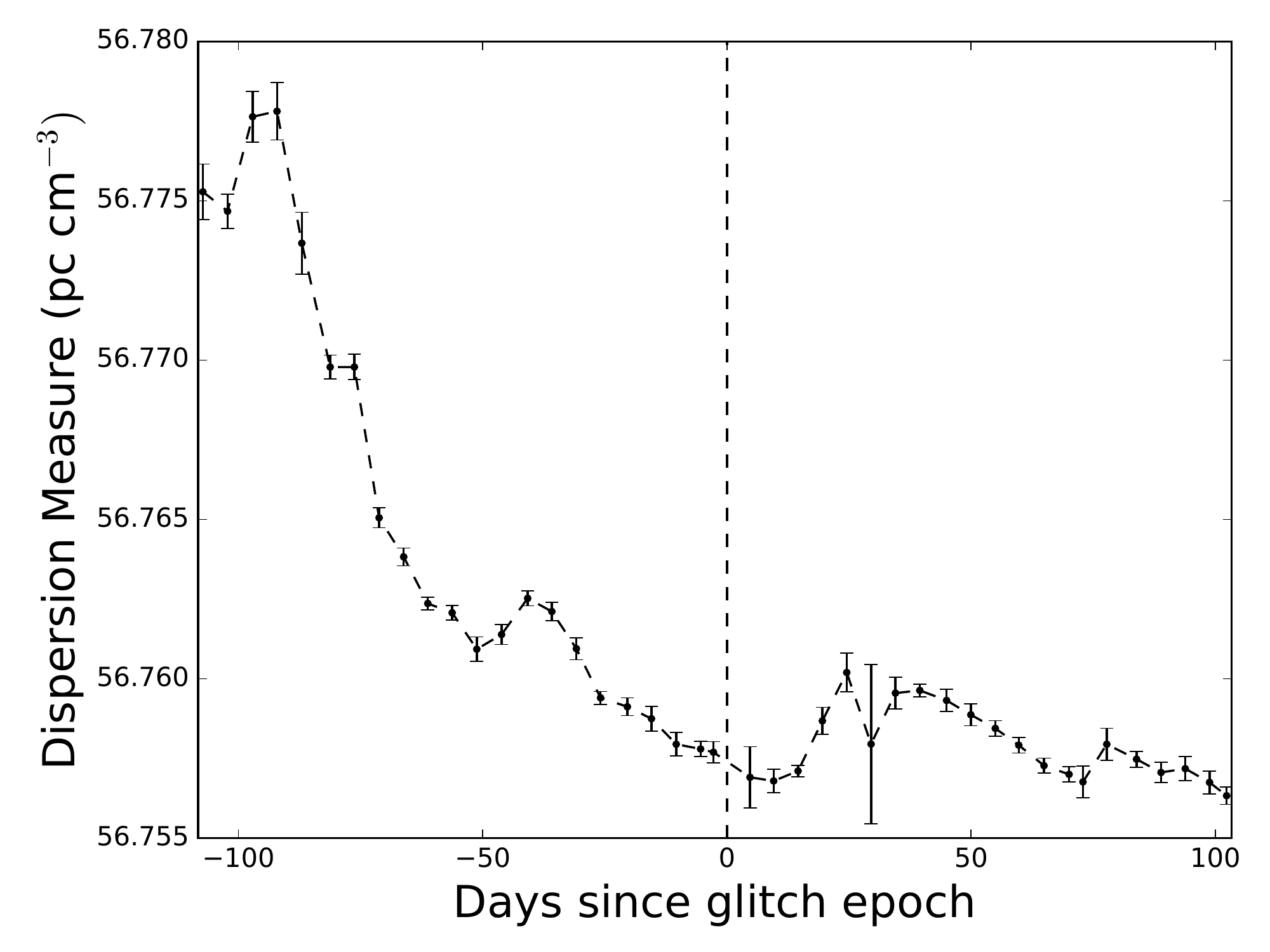}
    \caption{The evolution of the dispersion measure of the Crab pulsar over 200 days centred on the glitch epoch (dashed vertical line). }
    \label{DM}
\end{figure}

The Crab pulsar's dispersion measure (DM), is known to evolve in time due to the dynamic nature of the Crab Nebula (e.g., \citealt{glj11}). Therefore applying a constant value for the DM in the timing model can cause the integrated pulse profile to become broadened, sometimes causing a resultant shift in the computed arrival times. To examine the behaviour of the time variable DM in the Crab near the glitch, we employ the striding boxcar method describe above for 200 days of data centred on the glitch epoch. In each 10 day long segment, a fit is applied for rotational parameters up to second order and DM, and we stride over the data in steps of 5 days. The DM evolution over this period is shown in Figure \ref{DM}. In the initial $\sim$80 days, the DM undergoes a $\sim$1.5 per cent decrease which is typical of events seen in the Crab (e.g., \citealt{mck18}, submitted). Near the glitch, the DM is more slowly evolving, indicating that the measured rotational evolution near the glitch is not contaminated by variations in DM. These changes are well-modelled by a DM variation and can be corrected for.  

We also fit for the glitch parameters using \emph{TEMPO2}\footnote{\url{http://www.atnf.csiro.au/research/pulsar/tempo2/}} \citep{hem06}.  The fit includes rotational parameters up to second order and DM parameters up to first order.  We find the magnitude of the initial step to be $1.4233(5) \times 10^{-5}$~Hz and that of the delayed spin-up to be $1.071(4) \times 10^{-6}$ Hz. The delayed spin-up rises exponentially with a time constant of 1.703(13) days. The small increase in the spin-down rate is $-2.569(8) \times 10^{-12}$ Hz s\textsuperscript{-1} The pre-and post-glitch parameters are listed in Table \ref{glitch_table}.

\begin{figure}
\includegraphics[width=1.0\columnwidth]{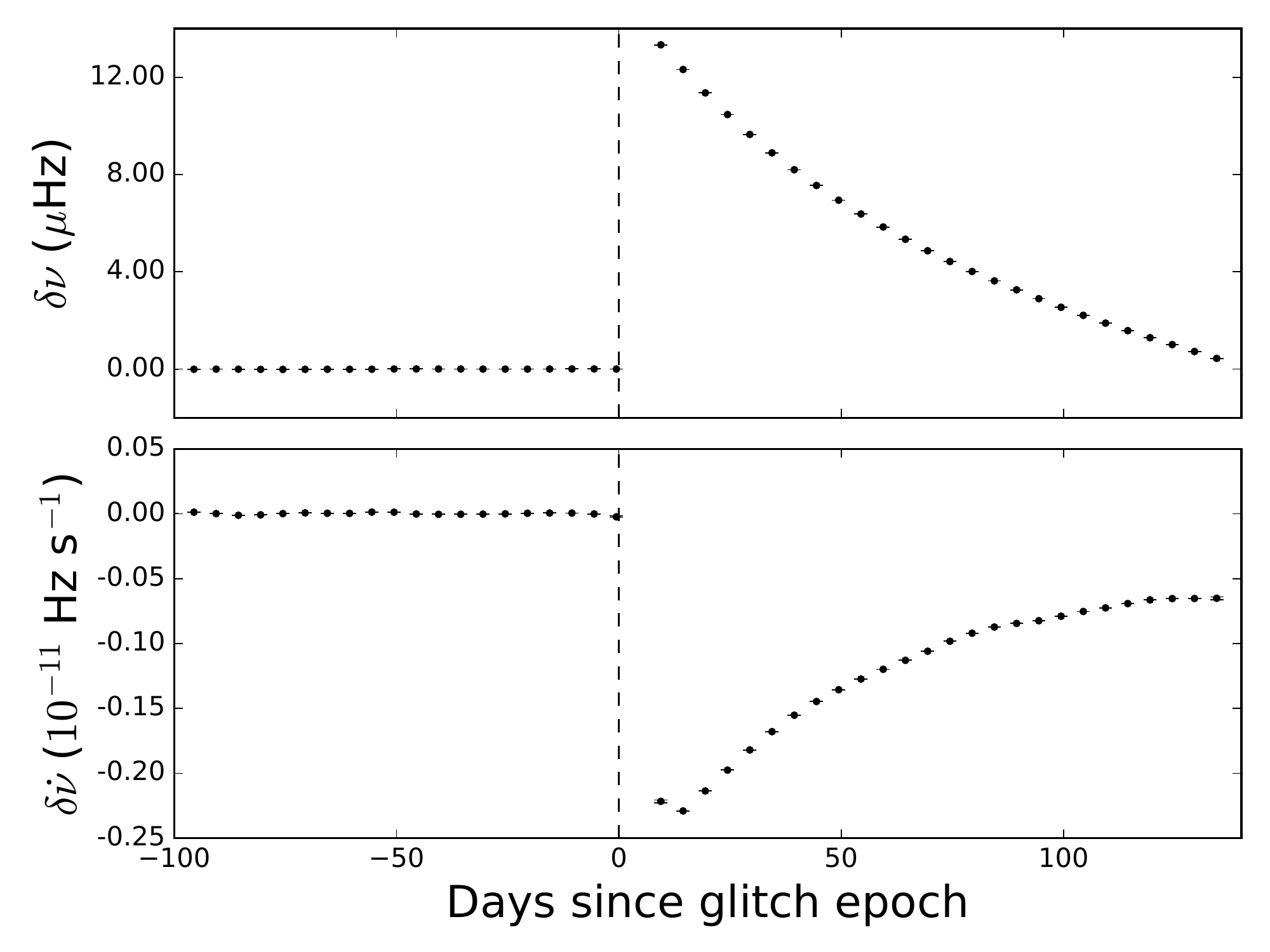}
   \caption{The medium term rotational evolution of the Crab pulsar near the glitch epoch (dashed line). Upper panel: The frequency residuals $\delta \nu$ relative to the extrapolation of the pre-glitch rate after subtracting the initial spin-down rate (Table \ref{glitch_table}). Lower panel: Frequency derivative residuals $\delta \dot{\nu}$ relative to the initial value. A downward inflection represents an increase to the magnitude of the spin-down.} 
   \label{nudot_lowres}
\end{figure}
    
In order to characterise the longer-term behaviour of the Crab pulsar's rotation, we repeat the boxcar stride using lower time resolution data. In this case, the TOAs from each 30 minute-long observation are combined into a single TOA, resulting in 2 TOAs per day allowing us to trade cadence for timing precision, averaging over the short-term transient behaviour very close to the glitch whilst still allowing us to carefully examine the slower recovery.  We use data from 100 days prior to the glitch to $\sim$150 days following the glitch.  We set a boxcar width of 20 days, striding in 5 day steps, fitting for values of $\nu$, $\dot{\nu}$ and $\ddot{\nu}$ at each stage. The result is shown in Figure \ref{nudot_lowres}. The transient behaviour  of the rotational parameters near the glitch are now unresolved.
The evolution of $\dot{\nu}$  (lower panel) is consistent with that of other large Crab glitches. Following the initial rapid rise, the value of $\dot{\nu}$ begins rapidly decreasing towards the pre-glitch value. In the glitches of 1989, 1996 and 2004, this partial recovery took $\sim$150 days after which it underwent a further turnover, increasing in value again over a long-term period of recovery lasting $\sim$3 years \citep{ljg+15}.  This results in a permanent increase in $\dot{\nu}$. At the time of writing, this second turnover into long-term recovery is yet to occur, with the value of $\dot{\nu}$ continuing to decrease towards the pre-glitch value. 

\subsection{Glitch-associated radiative changes}
\subsubsection{The integrated radio pulse profile}

\begin{figure}
    \includegraphics[width=\columnwidth]{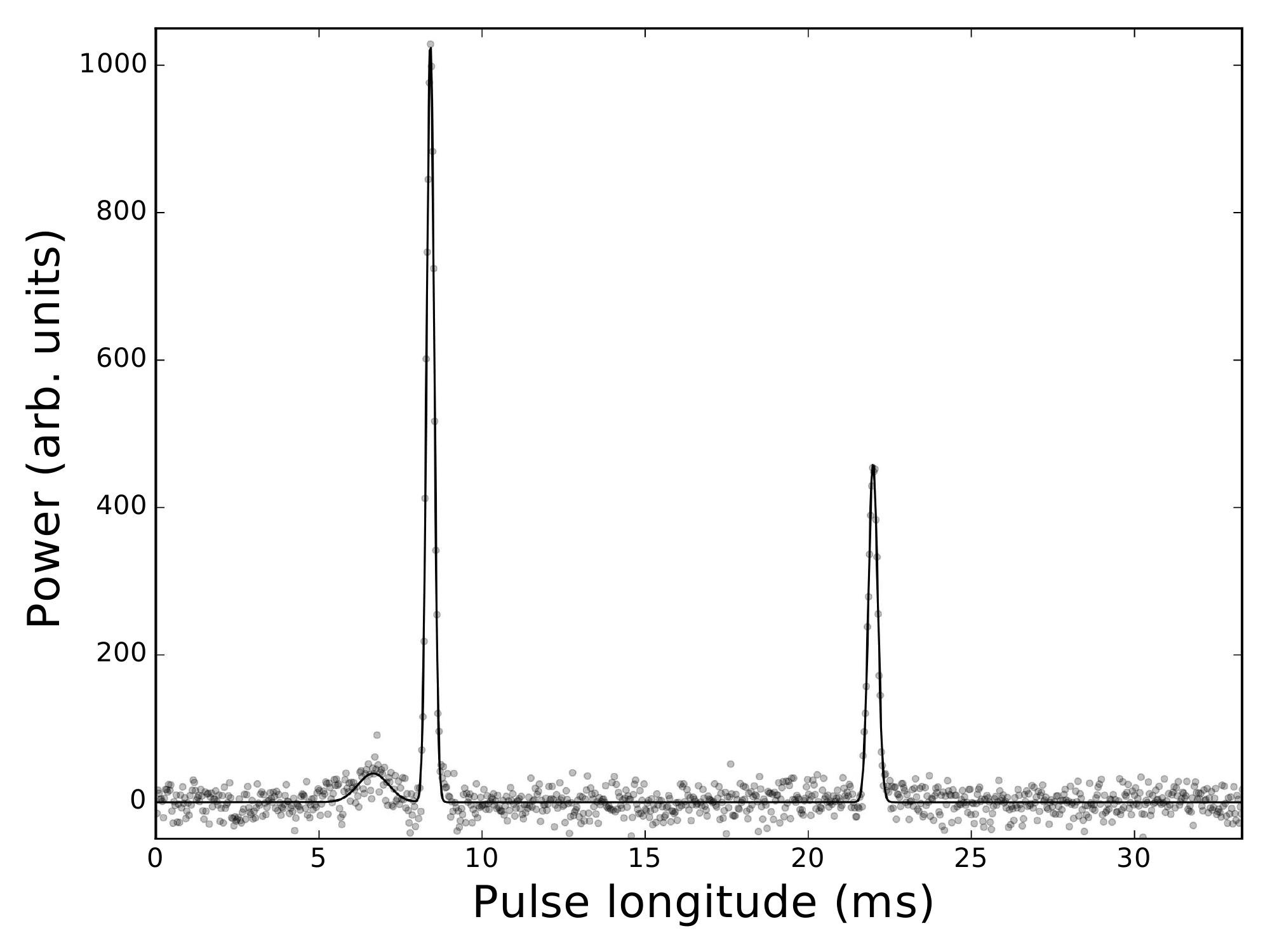}
    \caption{The integrated pulse profile of the Crab pulsar at 610 MHz. The profile over one full rotation of the star is shown, comprising 1024 bins corresponding to a time resolution of 0.0330 ms. The profile is formed from the aggregation of $\sim 10^6$ individual pulses observed over 11.5 hours on MJD 58056. The grey dots are the observed integrated profile. The black line corresponds to three Gaussian functions used to model the individual components (see text).}
    \label{template_crab}
\end{figure}

Figure \ref{template_crab} shows the integrated pulse profile of the Crab pulsar at 610 MHz. The profile is composed of three main components. The main pulse (MP) and the interpulse (IP) occurs at pulse longitudes of $\sim$8 ms and $\sim$22 ms respectively  (Figure \ref{template_crab}).  Leading the main pulse by $\sim$1.5 ms is a precursor component (PC) that decreases in power with increasing observing frequency.  At 1.4 GHz this component is very weak or not detected.  An additional, fourth weak component leads the PC and is seen in both 610 MHz and and 1.4 GHz observations, however this only becomes visible when many profiles are summed \citep{lgw+13}.  

\begin{figure*}
    \includegraphics[width=1.0\columnwidth]{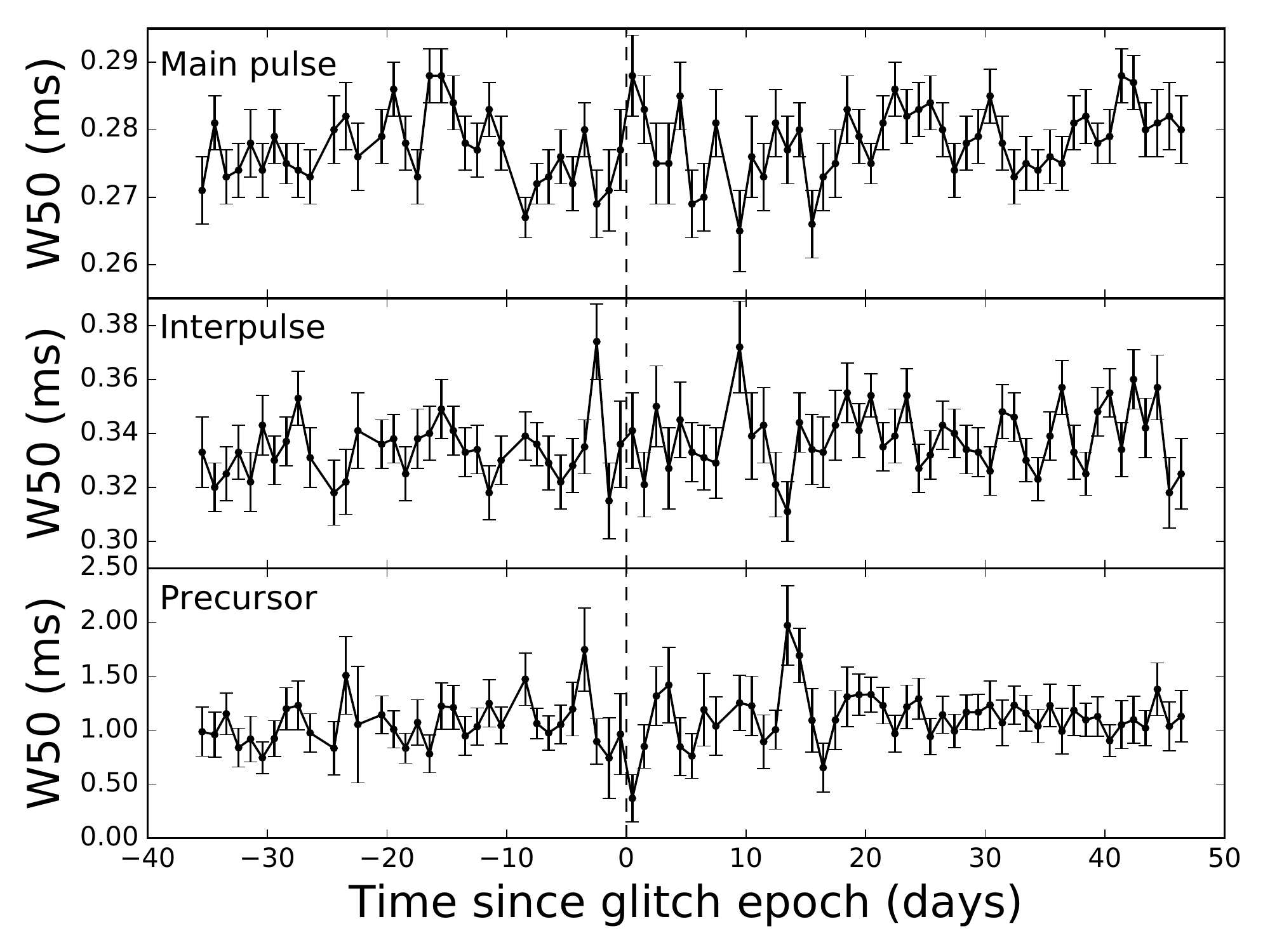}
    \includegraphics[width=1.0\columnwidth]{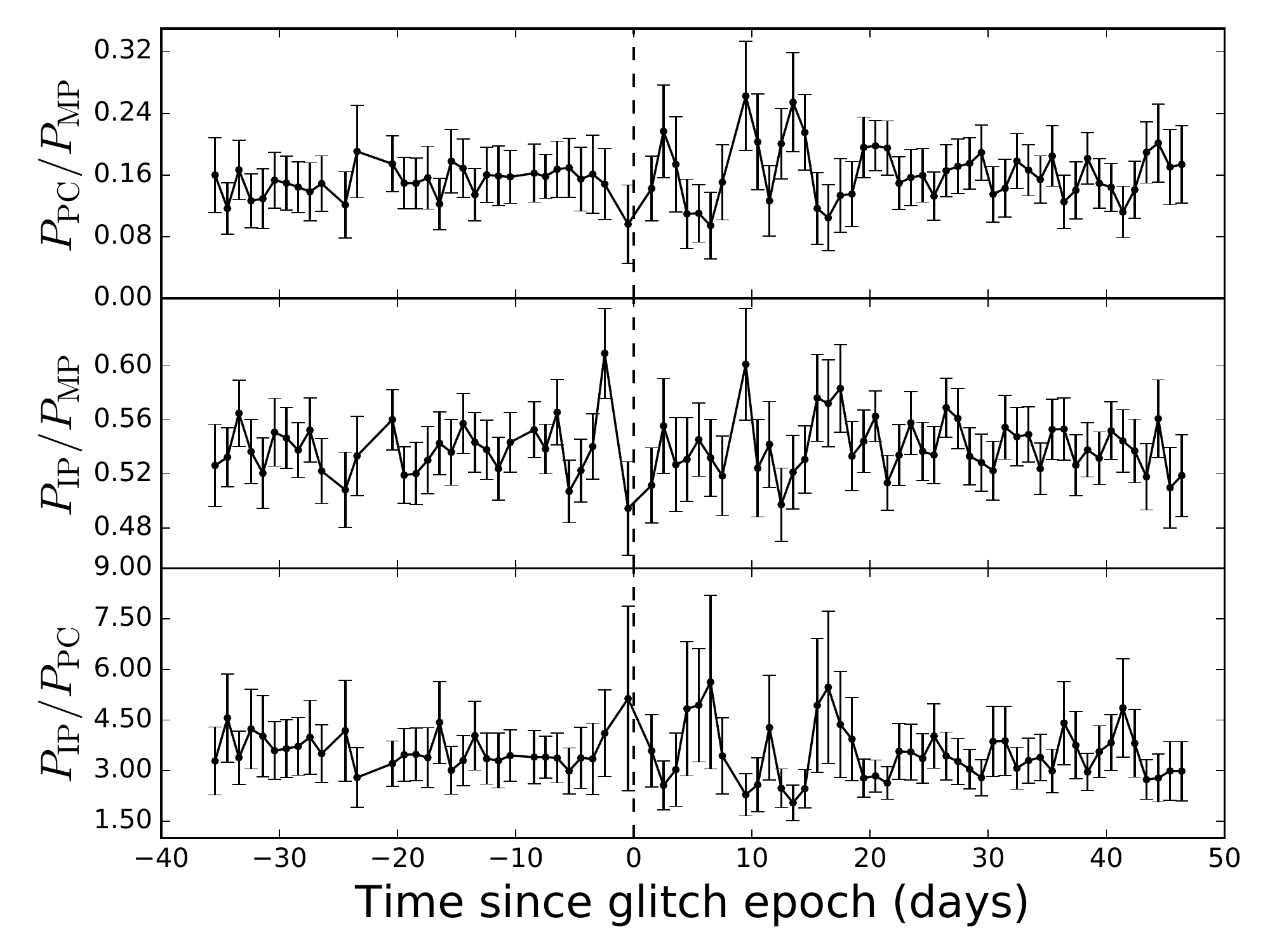} \\
    \includegraphics[width=1.0\columnwidth]{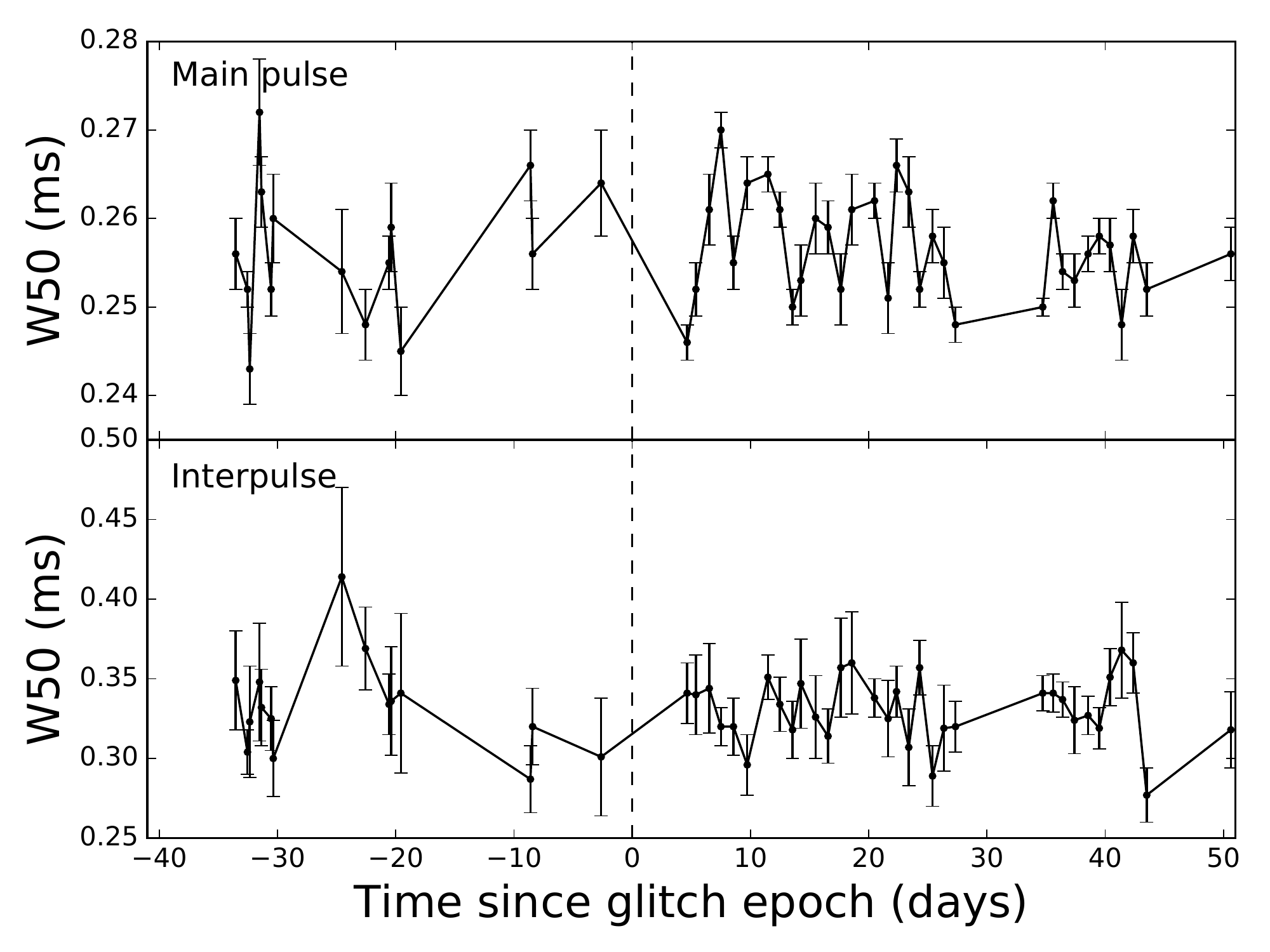} 
    \includegraphics[width=1.0\columnwidth]{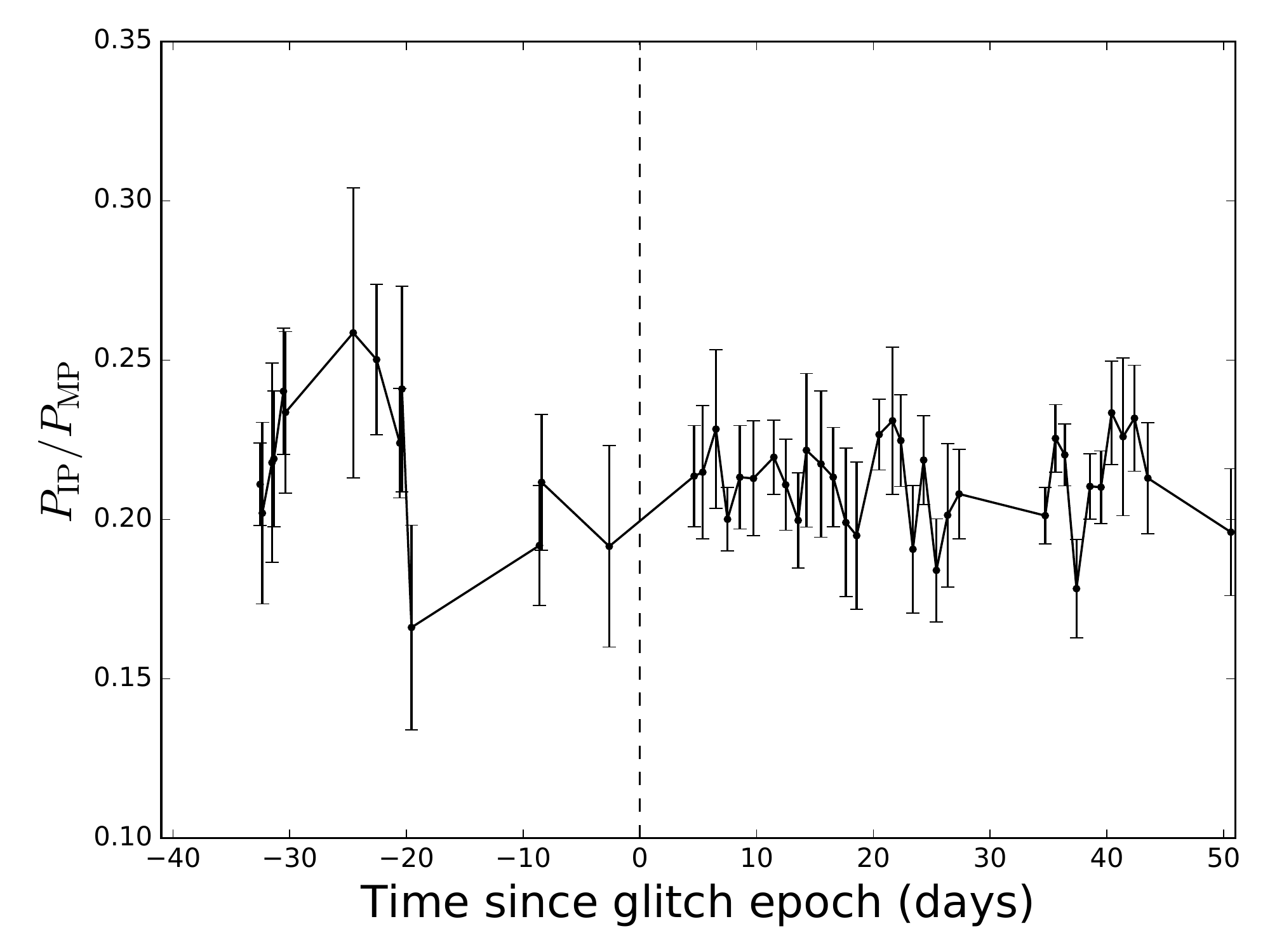} 

    \caption{Left panels: the evolution of the the full-width at half-power (W50) of each pulse profile component observed at 610 MHz (upper) and 1520 MHz (lower) over $\sim$ 90 days. The glitch epoch is denoted by the dashed line. Each data point represents the observations taken over a single day. The properties of the pulse components are measured by fitting a set of Gaussian functions to each of the integrated pulse profiles (see text). The error bars are obtained from the fit parameters for each component. Right panel: The evolution of the total power ratios for each combination of components for the 610 MHz (upper) and 1520 MHz data (lower). Only the MP and IP components are included for the 1520 MHz data. As left panel otherwise.}
    \label{pulse_evolution} 
\end{figure*}

Changes to the morphology of the radio pulse profile are not normally associated with glitch events.  If the glitch is triggered by an alteration in the conditions of the neutron star interior, it is not clear how this could induce changes in the configuration of the external magnetic field or the rate of pair creation, thereby leading to variability in the pulse profile. Such behaviour has been observed however, in two radio pulsars. The young, isolated, high-magnetic-field pulsar PSR J1119$-$6127 has undergone two large glitches with which coincident changes to the radio emission and pulse shape were seen (\citealt{wje11}; \citealt{awe+15}; \citealt{akts17}).{More recently, the Vela pulsar was shown to exhibit dramatic changes to the pulse profile and arrival times during the 2016 December glitch \citep{pdh+18}.} Using daily observation of the Crab pulsar with the OOTY radio telescope at 300 MHz, \cite{kjbm17} noted that the ratios of powers between each pulse component changed after the glitch. Specifically MP/IP, MP/PC and IP/PC all reduced respectively by factors of 1.2, 1.3 and 1.1.  This together with the large magnitude of $\Delta \nu$ prompts us to investigate the possibility of glitch-associated radio emission changes.

Pulse profiles are obtained for each individual 610 MHz and 1520 MHz observation from MJD 58016 to MJD 58083. For the 610 MHz data, where an observation begins less than 8 hours later than the previous one ended, the two observations are added together resulting in $\sim$1 observation per day over this period. 

Gaussian functions are fitted the to individual pulse profiles in order to model the individual profile components according to the procedure outlined in \cite{lgw+13}. The fit parameters for each Gaussian (the components' amplitudes, widths and positions) are used to quantify the evolution of the entire pulse over the MJD range of interest.  Figure \ref{pulse_evolution} shows the evolution of the widths at 50 per cent of the amplitude (W50) for each component and the ratios of the powers of each component in the 610 MHz data (upper plots) and the 1520 MHz data (lower plots).  The power in each component is evaluated by computing the product of its W50 and its amplitude. Due to its weakness, in the 1520 MHz data, we do not include the values for the PC component. Each data point represents one day of observations with the associated 1$\sigma$ errorbars obtained from the Gaussian fit parameters. There are no significant differences between the component widths before and after the glitch in neither the 610 MHz or the 1520 MHz data (left panels), nor are there any indications of transient changes to the widths near the glitch epoch. Similar stability is seen when considering the ratios of the total power under each component (right panels).  In addition, we average 7 days of 610 MHz data either side of the glitch and compare the power ratios, finding no significant differences between the two profiles. Therefore we find no evidence to suggest that the structure of the pulse profile was affected near the glitch at 610 MHz or 1520 MHz.   

\subsubsection{X-ray emission}

\begin{figure*}
    \includegraphics[width=1.0\columnwidth]{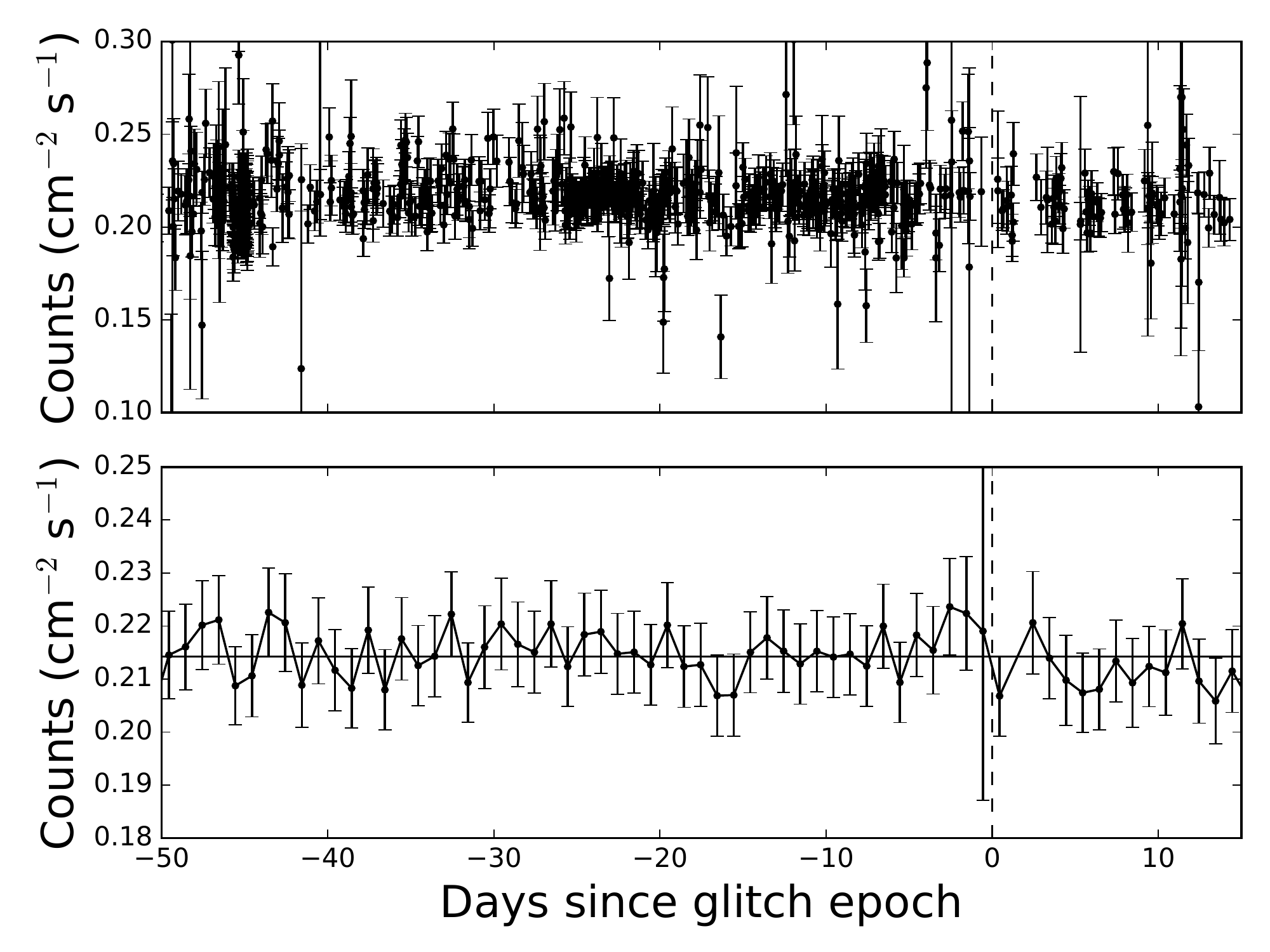} 
    \includegraphics[width=1.0\columnwidth]{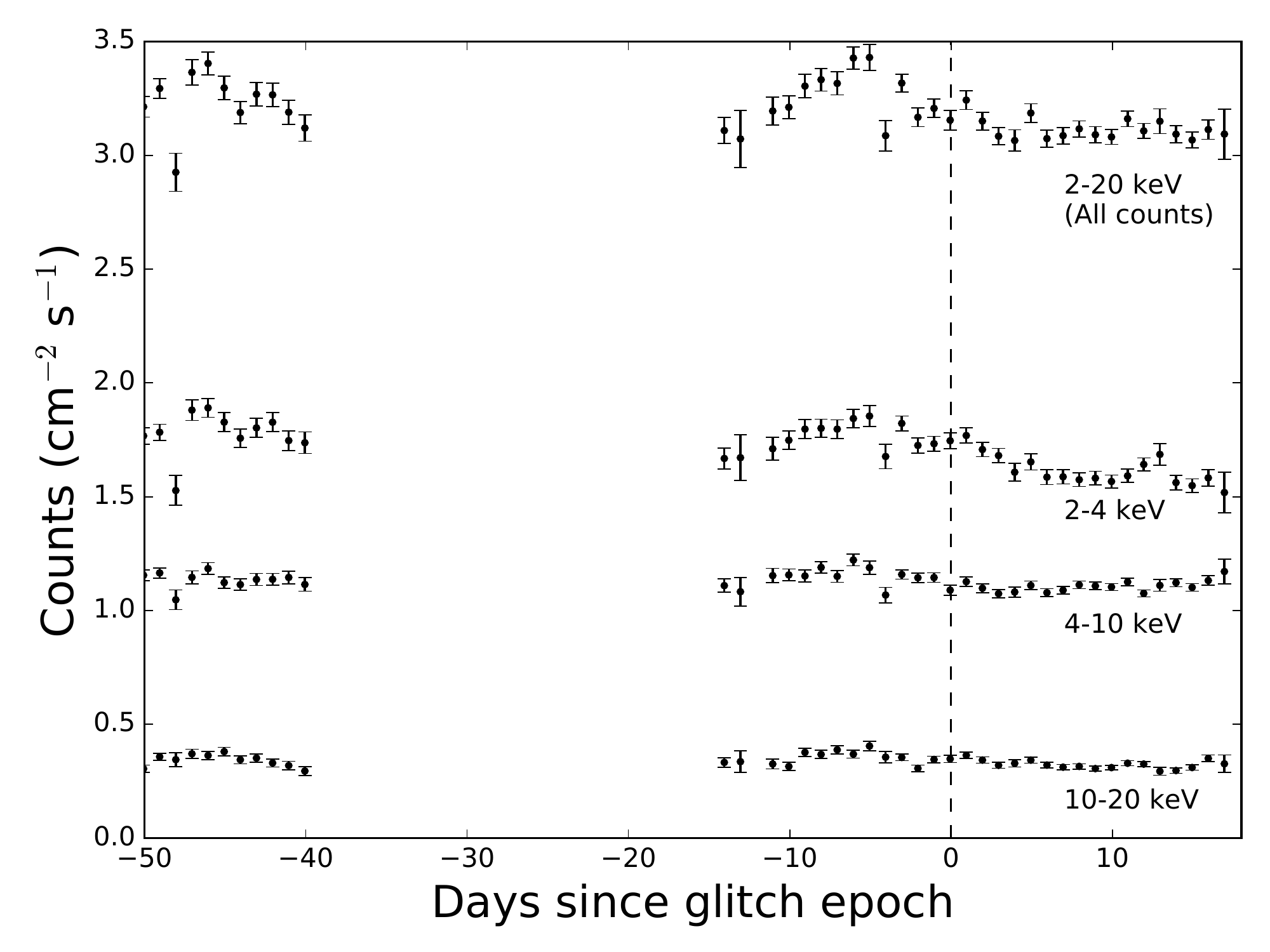}
    \caption{Left panels: The X-ray light curve of the Crab pulsar in the energy band 15-150 keV observed with the \emph{Swift}-BAT hard X-ray transient monitor. Fluxes are shown from 50 days prior to the November 2017 glitch until $\sim$15 days afterwards.  In the upper plot, each point is the average flux over the duration of a single pointing of Swift. Each point in the lower plot represents the weighted mean of all fluxes obtained in a single day. The horizontal line is the average daily flux (0.214 counts/cm\textsuperscript{2}/s) over the 65 days shown. Right panels: Daily X-ray fluxes over the same timescale for the individual energy bands, as well as the total counts observed with the MAXI instrument. Data were obtained from http://maxi.riken.jp on 1 March 2018. No data were available from between 40 and 15 days prior to the glitch.  The dashed vertical lines represent the glitch epoch.}
    \label{swiftprefix}
\end{figure*}

As young (though typically more strongly magnetised) pulsars have been shown to undergo high-energy radiative changes near large glitches (e.g., \citealt{gkw02}; \citealt{wje11}; \citealt{awe+15}), we examine the light curve of the Crab pulsar in X-rays to search for similar events close in time to the November 2017 glitch.

The Burst Alert Telescope (BAT) on-board the \emph{Swift} satellite \citep{gcg+04} routinely monitors the Crab. Light curve data in the energy band 15-50 keV are available from February 2005. These light curves are produced by the \emph{Swift}-BAT hard X-ray transient monitor, which utilizes the ``scaled-map data'' collected onboard as part of the process that searches for GRB triggers (see more details in \citealt{khc+13}).

Light curve data from the BAT are produced soon after observations are performed. As \emph{Swift} completes several pointings of the Crab per day, average fluxes are produced for each individual observation (the \emph{orbital} light curve). Additionally, a daily average flux is produced by computing the weighted mean of all fluxes measured over a given day. The orbital and daily light curves for the Crab pulsar, close in time to the November 2017 glitch are shown in Figure \ref{swiftprefix} (left panels).  


There are no obvious changes in the BAT light curve (15-50 keV) associated with the glitch. We note that a large number of the BAT scaled maps that are collected around Nov. 7, 2017 suffer from the so-called ``overflow problem'' due to the extreme outburst of the Be/X-ray binary Swift J0243.6+6124. This ``data overflow'' problem occurs when the collected counts exceed the maximum value that can be stored in the scaled map. As a result, the counts saturated with the extra counts are wrapped around and stored as extremely low values. These low values are thus not real and are removed \footnote{We note that as of Mar. 1, 2017, the BAT light curves on the transient monitor page \citep{khc+13} contain some data points with unusually low counts, including a dip around Nov. 7, 2017. After investigating each data point, we conclude that all of these low counts are due to the overflow problem and thus are not real.The BAT team is implementing a fix in the script to removes these data points that have the overflow problem. The online light curves will be updated in the near future.}. 

We also examine the Crab light-curve data from the Monitor of All-sky X-ray Image (MAXI) on board the International Space Station \citep{mku+09}. MAXI produces daily light curves in the energy bands 2-20 keV, 4-10 keV and 10-20 keV as well as in the total band 4-20 keV. There are no significant changes to the X-ray flux in any of the MAXI energy bands (Figure \ref{swiftprefix}, right panel). We therefore do not identify any changes to the 2-50 keV X-ray flux from the Crab pulsar that are associated with the November 2017 glitch.


\section{Discussion}

\begin{figure}
   \includegraphics[width=\columnwidth]{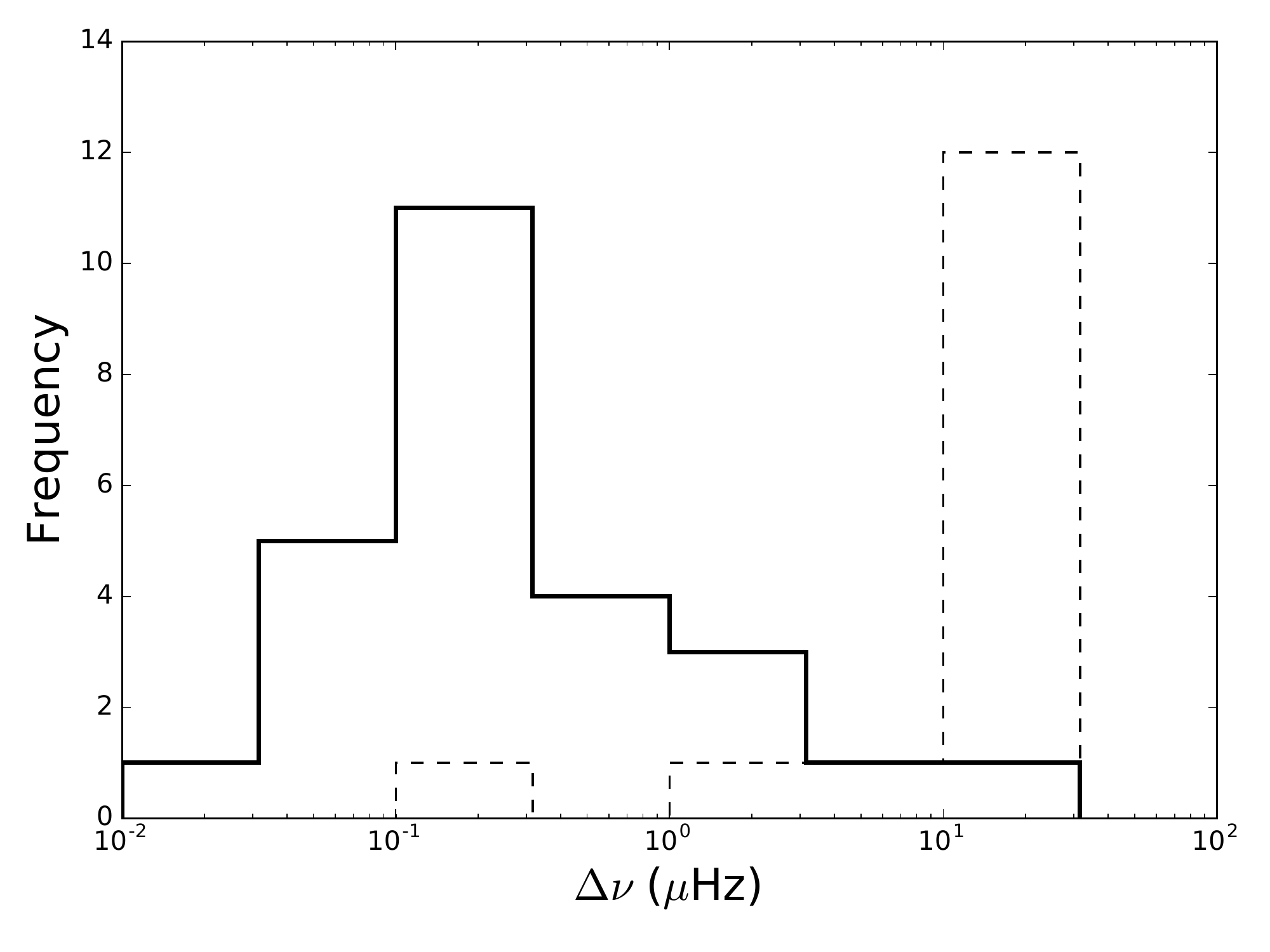}
   \caption{Histogram of glitch amplitudes ($\Delta \nu$) for the Crab (solid) and Vela (dashed) pulsars since monitoring began in 1968.}
   \label{crabvelahist}
\end{figure}

The observed rotational history of the Crab pulsar now includes 25 glitches.  The most recent glitch is the largest ever recorded in this source with a fractional, partially resolved spin-up $\Delta \nu / \nu \sim 0.516 \times 10^{-6}$. This is more than twice the size of the previous largest which occurred in March 2004. The glitch amplitude of the 2017 event is comparable to those seen in the Vela pulsar (see Figure \ref{crabvelahist}), whose glitches are consistently large with $\Delta \nu / \nu \sim 10^{-6}$ \citep{ymhj+13}.  Though the onset of the glitch occurred whilst the pulsar was not being observed, our long dwell-time and daily monitoring of the Crab has allowed us to identify the glitch just hours after it commenced, as well as resolve very short-term transient behaviour around the glitch epoch, including a delayed spin-up of the pulsar during the glitch itself. 

In previous large glitches (that are relatively isolated in time), the Crab has demonstrated a unique behaviour in terms of how the long-term recovery of $\dot{\nu}$ progresses. The value of $\dot{\nu}$ undergoes a rapid increase in magnitude (i.e., becoming more negative), typical of other young glitching pulsars, before rapidly declining in magnitude over several months, but only part-way towards the pre-glitch value. Following this, the magnitude of $\dot{\nu}$ rises exponentially again over a timescale of $\sim$3 years. This results in large glitches leaving a long term persistent offset in the value of $\dot{\nu}$ over time (see \cite{ljg+15}, Figure 3). At the time of writing, insufficient time has elapsed since the 2017 glitch to truly understand the long term recovery. Nevertheless, the magnitude of $\dot{\nu}$ is undergoing a rapid decline, following its initial increase, that is typical of previous Crab events (see Figure \ref{nudot_lowres}). 

\cite{ljg+15} demonstrated, using 15 Crab glitches, that the value of the persistent offset $\Delta \dot{\nu}_\mathrm{p}$ is loosely dependent on $\Delta \nu$ via the relation (see Figure 5 therein),

\begin{equation}
    |\Delta \dot{\nu}_\mathrm{p}| = 7 \times 10^{-8} \times \Delta \nu.
\end{equation}

\noindent If this relation is applicable also to this glitch then after approximately three years we expect the persistent offset to have a value of $\Delta \dot{\nu}_\mathrm{p} \sim 1 \times 10^{-12}$ Hz s\textsuperscript{-1} corresponding to a permanent fractional increase in $\dot{\nu}$, due to the glitch, of $\sim$3 per cent, provided that no subsequent glitches occur during that time.

\subsection{The Crab glitch size distribution} \label{sizes}

In Figure \ref{crabvelahist}, we show the difference in glitch amplitudes between the Crab and Vela pulsars as the rotation of these frequently glitching sources has historically been the focus of much study.  Prior to the latest Crab glitch, Vela (dashed line) has almost consistently exhibited larger ($\Delta \nu \sim$10\textsuperscript{-5} Hz) , more narrowly distributed glitches than the Crab (solid line) since observations of both sources began. The Crab undergoes a much wider distribution of glitch sizes, spanning four orders of magnitude. The smallest Crab glitch to date had a $\Delta \nu = 5 \times 10^{-8}$ Hz - roughly an order of magnitude smaller than the smallest glitch in Vela. 

The lack of large Crab glitches has been attributed to the relative youth of the pulsar (e.g., \citealt{wwty12}).  Younger pulsars are expected to have a greater oblateness due to their higher spin-rates (e.g., \citealt{r69}; \citealt{ml90}). Older pulsar surfaces are cooler, resulting in stresses building up in the surface due to spin-down. Relief from this stress, in the form of \emph{crustquakes}, has the effect of suddenly reducing the moment of inertia of the star, thereby increasing the rotation rate. In younger pulsars whose surfaces have a higher temperature, the surface stresses are less able to accumulate due to greater surface plasticity, resulting in smaller glitches. The fact that the Crab has exhibited a glitch comparable in magnitude to the Vela pulsar ($\tau_{\mathrm{char}}\sim11,000$ years), and that large glitches have been observed in hot young pulsars that also undergo magnetar-like X-ray bursts (e.g., \citealt{wje11}; \citealt{awe+15}; \citealt{akts17}) is inconsistent with such a model.

\begin{figure}
   \includegraphics[width=\columnwidth]{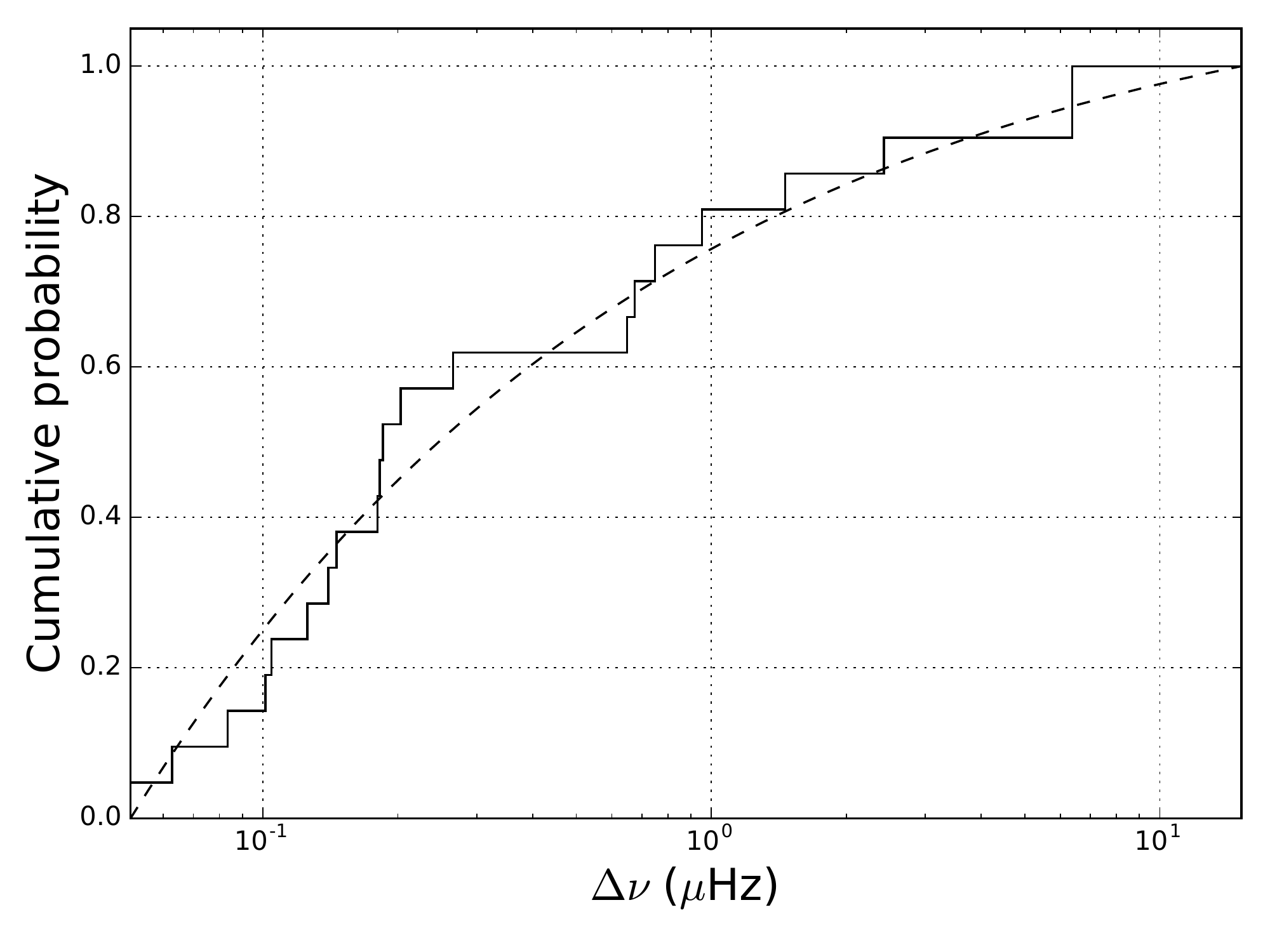}
   \caption{The cumulative distribution of the glitch amplitudes observed in the Crab pulsar. The distribution includes 21 glitches observed since 1984.  The dashed line is the power law fit to these data given by Equation \ref{powerlaw} (see Section \ref{sizes}).}
   \label{cdf}
\end{figure}

\cite{eas+14} demonstrated that the distribution of glitch sizes in the Crab can be adequately described by a power law of the form,

\begin{equation}
    p(\Delta \nu) = \frac{(1 - \alpha) \Delta \nu^{-\alpha}}{\Delta \nu_{\mathrm{max}}^{(1 - \alpha)} - \Delta \nu_{\mathrm{min}}^{(1 - \alpha)}}, 
    \label{powerlaw}
\end{equation}

\noindent where $\Delta \nu_{\mathrm{min}}$ and $\Delta \nu_{\mathrm{max}}$ are the minimum and maximum glitch sizes, respectively, in the sample. The power law index was calculated to be $\alpha = 1.36$. In order to compute whether the glitch size distribution remains consistent with this power law, we recalculate $\alpha$ using an updated sample that includes the November 2017 glitch. To do this we employ the maximum likelihood method to estimate the most likely value for $\alpha$. Glitches that occurred prior to 1984, when daily monitoring began, are not included as the sample up to this date may not be complete. We also exclude the February 1997 (MJD 50489) glitch as its classification as a glitch is uncertain \citep{wbl01} due to the timing residuals near this event being more characteristic of timing noise. We note that this event was also not included in the power law analysis in \cite{eas+14}.  With $\Delta \nu_{\mathrm{min}} = 0.05 \mu$Hz and $\Delta \nu_{\mathrm{max}} = 15.21 \mu$Hz, we find $\alpha = 1.36 \pm 0.11$. This value is consistent with that calculated in \cite{eas+14}, showing that the November 2017 event could arise from the same underlying distribution. Figure \ref{cdf} shows the cumulative distribution function of post-1984 glitch sizes in the Crab and the associated power law fit (dashed line). In order to quantitatively test the hypothesis that the measured glitch amplitudes are drawn from the power law distribution we compute the Kolomogorov$-$Smirnov (KS) test statistic $D = 0.1$ and its associated $P$-value, $P = 0.9$. There is no evidence therefore that the observed sample is not drawn from the model distribution.   

The fact that glitches in some individual pulsars follow a power law distribution of sizes is consistent with the hypothesis that avalanche processes underpin their occurrence \citep{mpw08}. In such a process, the behaviour of individual components (in this case the superfluid vortices) is dominated by local \emph{near neighbour} forces rather than long timescale external driving forces (e.g., the growing lag between the crust and fluid interior velocities).  In this model, a fraction of the vortices in the fluid interior are pinned to defects or lattice sites in the inner crust. As the rotating superfluid slows, free vortices move outward and are lost at the edges of the fluid, creating a vortex density that is radially inhomogeneous. This causes a local increase to the forces on pinned vortices which, above some threshold, overcomes the pinning force and causes the vortex to unpin. As this unpinning causes further perturbations to the velocity of superfluid, other nearby loosely pinned vortices become unstuck from their lattice sites, triggering an avalanche.  As these newly freed vortices are now able to move outwards towards the inner crust, the rotational velocity of the crust increases to compensate.  If avalanche processes are responsible for pulsar glitches then a power law of the form of Equation \ref{powerlaw} should describe the glitch size distribution for individual pulsars. The Crab glitch size distribution has been previously shown to conform to such a power law (e.g., \cite{mpw08}; \cite{eas+14}) and the large 2017 glitch does not result in a departure from the Avalanche model.  Avalanche processes do not conform to any particular scale, allowing a wide range of glitch amplitudes such as those seen in the Crab. Conversely, the Vela pulsar's narrowly distributed and quasi-periodic glitching behaviour is symptomatic of a process which is dominated by global (e.g., the fluid-crust lag) rather than local near-neighbour forces. 

{Although within the bounds of the largest and smallest Crab glitches, the size distribution follows a power law relation, the substantial minimum glitch size (corresponding to the power-law cut-off at low $\Delta \nu$) demonstrated by \cite{eas+14} does not truly meet with the `scale invariance' requirement of systems driven by avalanche dynamics.  However, \cite{has16} showed that the effects of superfluid dynamics are consistent with the existence of a minimum glitch size, whilst vortex unpinning events can still obey scale invariance.}

\subsection{Glitch waiting times}

\begin{figure*}
   \includegraphics[width=1.02\columnwidth]{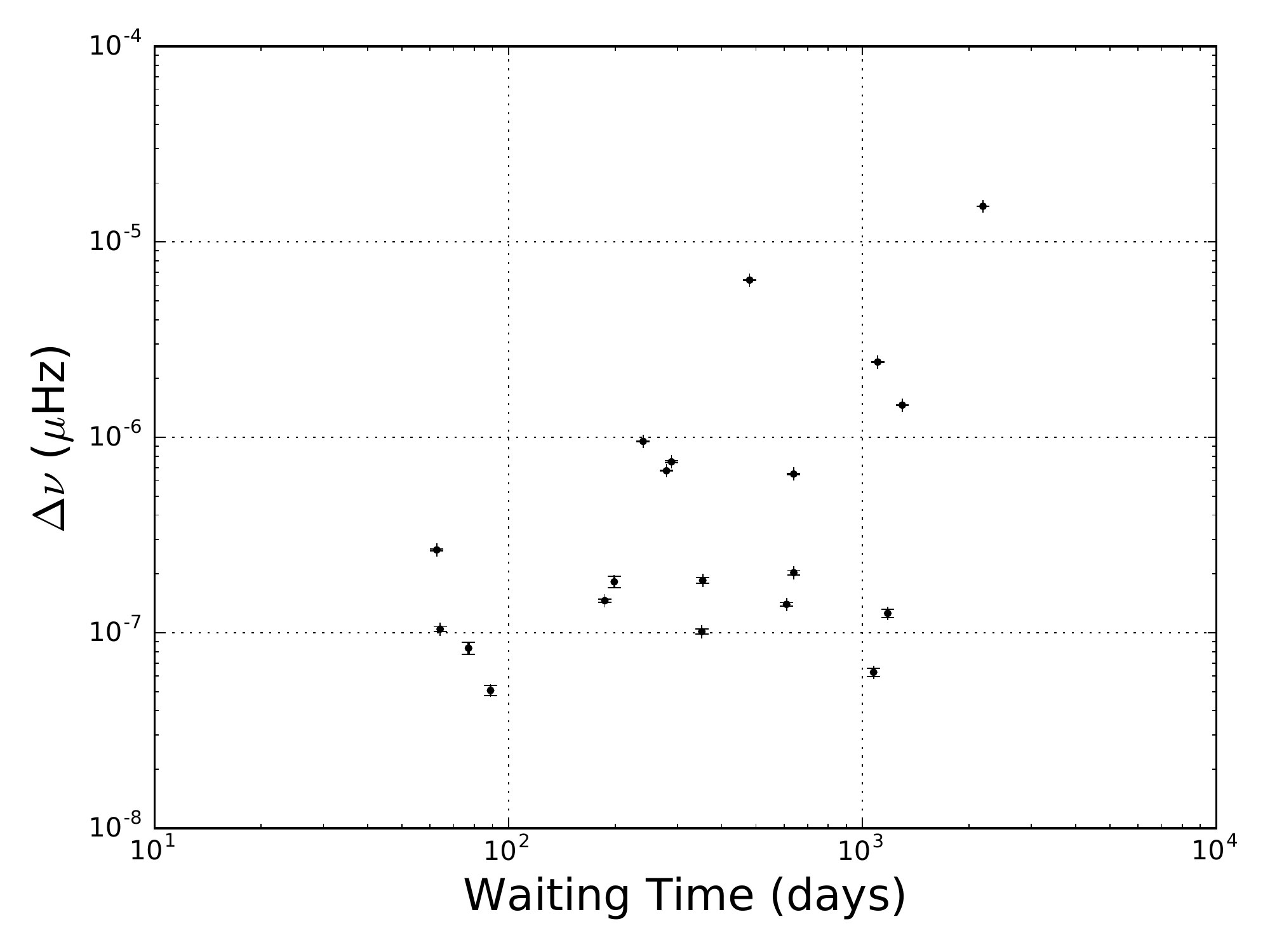}
   \includegraphics[width=\columnwidth]{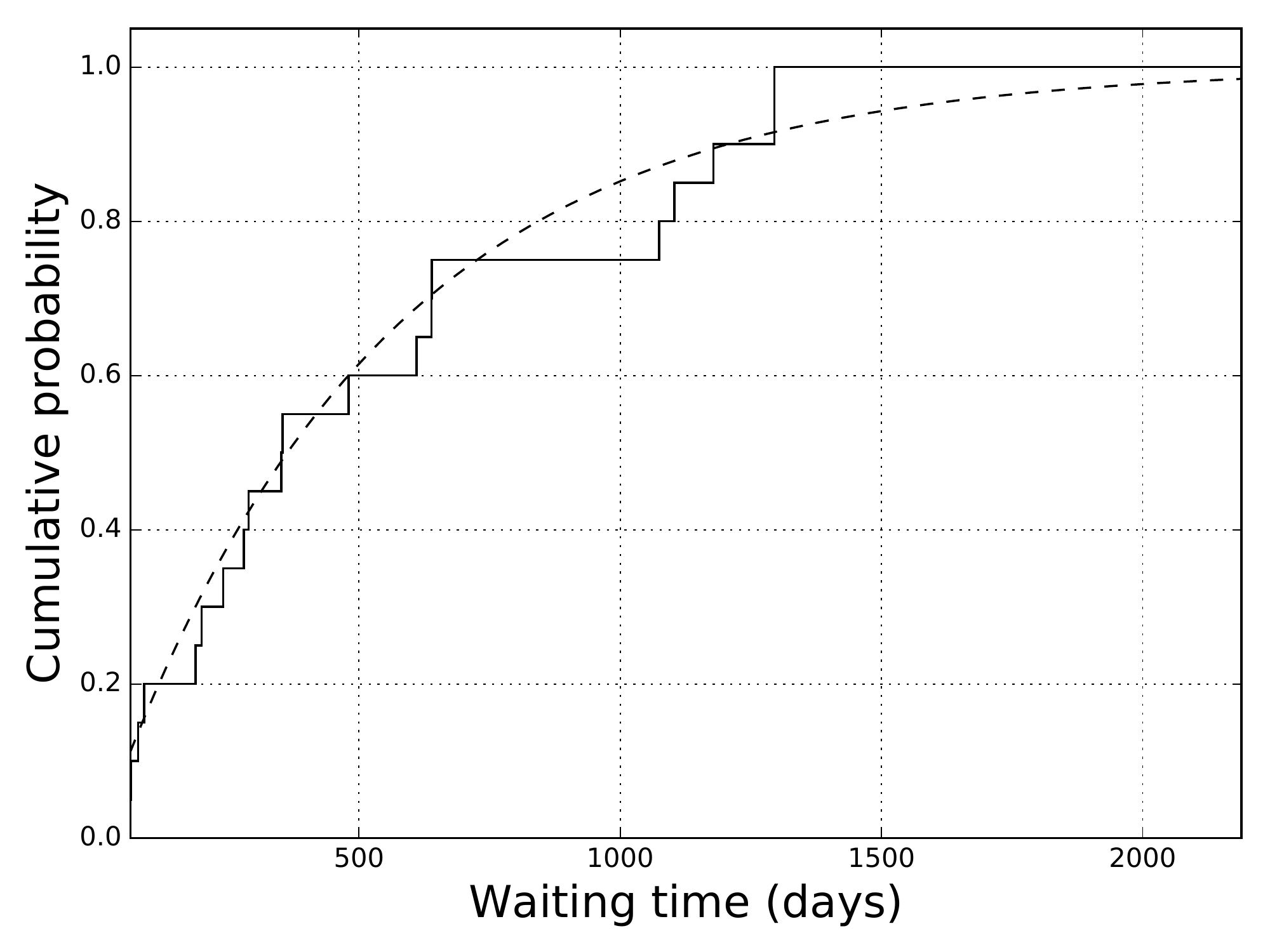}
   \caption{The relation between waiting time and glitch size for the 21 Crab pulsar glitches since 1984. The waiting time of a glitch is defined as the interval between the glitch and the previous glitch. This corresponds to 20 waiting times. The 2017 event is the upper-rightmost point.}
   \label{wtime}
\end{figure*}

In addition to being the largest glitch observed in the Crab pulsar, the 2017 event occurred after the longest period of glitch inactivity (2189 days) since 1984.  Figure \ref{wtime} (left panel) shows the size of all post-1984 Crab glitches against the time since their preceding glitches (defined here as the \emph{waiting time} $\Delta T$). Though a weak correlation is apparent, we note that this is strongly dominated by the 2017 event and a small number of small glitches have occurred after long waiting times.  Nevertheless, it is interesting that no large glitches have occurred after short intervals of inactivity. We compute the Spearman Rank correlation coefficient (SRCC) between $\Delta \nu$ and $\Delta T$ to be 0.40, indicating a moderate correlation between the two quantities.

In order to test the dependence of the SRCC on any one of the 20 members of the sample shown in Figure \ref{wtime}, we recalculate the SRCC 20 times, each time with one of the data points subtracted. We find that values of the SRCC range from 0.30 to 0.51. The lowest value of 0.30 occurs when the 2017 event is not included in the SRCC calculation, indicating that this event has a significant effect on the correlation. That said, in all cases, the correlation remains positive.


Despite the apparent correlation, given the small sample size, it remains possible that the observed distribution of sizes and waiting times has arisen by chance. To evaluate the probability of this we assume that the values of $\Delta \nu$ and $\Delta T$ are random and are each draws from log-uniform distributions.  We randomly draw 20 values of each quantity, bound by their minimum and maximum observed values, form 20 ordered pairs and calculate the SRCC. We repeat this 10\textsuperscript{7} times.  As expected, the distribution of resulting SRCCs is a Gaussian distribution with a mean value of zero. The standard deviation ($\sigma$) of SRCCs is 0.23.  The SRCC corresponding to the sample in Figure \ref{wtime} is $1.8\sigma$ away from the mean. We therefore conclude that the probability of the observed $\Delta \nu - \Delta T$ distribution arising by chance is 7 per cent. 

{A waiting time dependence where larger glitches occur after longer periods of glitch inactivity could be the consequence of a supply of stored angular momentum that becomes fully deposited into the crust at a glitch (the \emph{reservoir effect}). Conversely a correlation between the glitch amplitude and the time until the \emph{next} glitch would imply that glitches are triggered when a critical lag is reached between the superfluid and the crust. PSR J0537$-$6910 exhibits such a correlation very strongly (\citealt{aeka18}; \citealt{fagk18}) however we find no such correlation in the Crab pulsar. }

If glitches are statistically independent events (in other words, the probability of the occurrence of glitch $i$ is independent of the time since glitch $i-1$), then the distribution of waiting times should follow a Poissonian probability density function of the form

\begin{equation} \label{poisson}
    p(\lambda, t) = \lambda^{-1} \exp{(-t/\lambda)},
\end{equation}

\noindent where $\lambda$ is the mean waiting time. It has been demonstrated, prior to the 2017 event, that the distribution of waiting times for the Crab pulsar is well described by Poissonian statistics (e.g., \citealt{wbl01}; \citealt{mpw08}; \citealt{wwty12}). We show a cumulative distribution of glitch waiting times, updated to include the 2017 event, in Figure \ref{wtime} (right panel).  Fitting a Poisson model to the data yields a mean waiting time of $\lambda = 524$ days. Predictably, due to the long interval between the 2017 event and its predecessor, this is somewhat greater than the value computed in \cite{wwty12} ($\lambda = 419$ days), whose sample differed to ours by only the latter two Crab glitches. We compute the KS statistic between our data and the Poisson model to be $D = 0.1$ with an associated P-value, $P = 0.9$ indicating good agreement. 

\subsection{The partially resolved spin-up}

The initial unresolved spin-up amplitude of the November 2017 glitch was measured to be $\Delta \nu = 14.2$ $\mu$Hz. This constitutes approximately 93 per cent of the total spin-up amplitude. {We place an upper limit on the rise time of the unresolved component of $\sim$6 hours - corresponding to the time between the glitch epoch and the commencement of the first post-glitch observation.}  The remaining 7 per cent ($\Delta \nu_{d} = 1.1$ $\mu$Hz) was resolved in time over a period of $\tau_{d} = 1.7$ days.  In the Crab pulsar, this is the third reported instance of a partially resolved spin-up. The spin-ups of the 1989 and 1996 glitches were partially resolved in time over 0.8 days and 0.5 days respectively. These spin-up timescales are substantially greater than those predicted by fluid-crust coupling models (e.g., \citealt{als84}).  The Crab may be unique in this respect. Unfortunately, resolved spin-ups (partial or otherwise) have only ever been observed on these three occasions. {In contrast, the Vela pulsar's 2016 December glitch occurred whilst the pulsar was being observed  \citep{pdh+18} and the total spin-up ($\Delta \nu = 16.0$ $\mu$Hz) was seen to  occur in $\sim$5 seconds }

Delayed increases in the spin frequency have been attributed to a fraction of the unpinned vortices moving inwards towards the core \citep{accp96}, post-glitch impedance of the inward motions of vortices \citep{rzc98},{deposition of thermal energy into the crust (e.g., through starquakes) \citep{ll02} and most recently to the formation of vortex sheets \citep{kh18}.} We note that in the three occasions where extended spin-ups have been observed, the glitches were comparatively large ($\Delta \nu = 2.4$ $\mu$Hz for the 1989 glitch and $\Delta \nu = 1.0$ $\mu$Hz for the 1996 glitch).  No such behaviour has been observed to date in smaller glitches even though in some cases, observations began within hours of a glitch epoch supporting the hypothesis made in \cite{wbl01} that extended spin-ups are only a feature of larger Crab glitches. We also note that in the three Crab glitches in which extended spin-ups were observed, larger glitch amplitudes $\Delta \nu$ correspond to longer rise times $\tau_d$ however, many more large glitches would be required to verify whether any genuine correlation exists.

\section{Conclusions}

Using daily, long dwell-time observations of the Crab pulsar (PSR B0531+21) at 610 MHz and additional observations at 1520 MHz, we have observed and measured the largest glitch in the source since observations of it began in 1968. In November 2017, the pulsar underwent a total spin-up of $\Delta \nu = 1.53037(30) \times 10^{-5}$ Hz. This corresponds to a total fractional spin-up of $\Delta \nu / \nu = 0.51637(10) \times 10^{-6}$.  This makes the November 2017 event more than twice the size of the previous largest Crab glitch in March 2004. We also measured a change to the spin-down rate of $-2.569(8) \times 10^{-12}$ Hz s\textsuperscript{-1}.  We summarise our analysis below:

\begin{itemize}
 \setlength\itemsep{1em}
 \item{The  glitch of November 2017 is similar in size to those consistently exhibited by the Vela pulsar (PSR B0833$-$45). The sizes of Crab glitches are now known to span more than three orders of magnitude. The size distribution of the Crab pulsar's glitches remains consistent with a cut-off  power law distribution computed in previous studies (e.g., \citealt{eas+14}) with a power law index $\alpha = 1.36$. }
 
 \item{7 per cent of the total spin-up was resolved in time over a timescale of 1.7 days. Partially resolved spin-ups were also seen in the Crab glitches of 1989 and 1996.}
 
 \item{Though subtle changes to the amplitudes of the radio profile components were claimed at 327 MHz, within the sensitivity of the 42-ft and Lovell telescopes, no such changes are seen at 610 MHz or 1520 MHz.}
 
 \item{Using light curve data from the \emph{Swift}-BAT instrument, we find no glitch-associated changes to the X-ray flux from the Crab in the energy band 15-50 keV. The same is true in the energy band 2-20 keV observed with the MAXI instrument.} 

 \item{Though largely dominated by the 2017 glitch, a correlation between the glitch amplitude and the time since the last glitch (the waiting time) is apparent. Monte-carlo simulations show that there is a 7 per cent probability that this correlation is false. We also note an absence of large glitches after short periods of inactivity.}
 
 \item{At the time of writing, the pulsar continues to undergo medium term recovery from the glitch and this continues to be monitored daily. In addition to the rapid spin-up, a large initial increase in the magnitude of $\dot{\nu}$ is observed that is followed by a partial recovery \emph{towards} the pre-glitch $\dot{\nu}$ over a period of $\sim$100 days. Following this, in line with other large Crab glitches, we expect $\dot{\nu}$ to exponentially decrease again over a period of years, provided the long-term recovery is not interrupted by future glitches.}
\end{itemize}

\section*{Acknowledgements}

We thank the anonymous referee for their constructive comments and suggestions for improvement to this manuscript. We also thank Danai Antonopoulou, Crist\'{o}bal Espinoza, Laura Driessen and Mark Kennedy for useful discussions. This research has made use of the MAXI data provided by RIKEN, JAXA and the MAXI team.  Pulsar research at Jodrell Bank Centre for Astrophysics and Jodrell Bank Observatory is supported by a consolidated grant from the UK Science and Technology Facilities Council (STFC).




\bibliographystyle{mnras}
\bibliography{journals,psrrefs,modrefs} 




\appendix


\bsp	
\label{lastpage}
\end{document}